\documentclass[journal]{IEEEtran}

\usepackage{amsmath,amsfonts,amssymb}
\usepackage{comment}
\usepackage{graphicx,subfigure}
\usepackage{times}
\usepackage{bm,bbm}
\usepackage{amsmath,amssymb}
\usepackage{graphicx,subfigure}
\usepackage{url}
\usepackage{units}
\usepackage[binary-units]{siunitx}
\usepackage{cite,float}
\usepackage{booktabs}
\usepackage{microtype}
\usepackage{euscript}
\usepackage{color,colortbl}
\usepackage{multirow,tabularx}
\usepackage[affil-it]{authblk}
\usepackage[dvipsnames]{xcolor}
\usepackage[hidelinks,colorlinks,breaklinks,naturalnames,bookmarks=false]{hyperref}
\usepackage{tikz}
\usetikzlibrary{calc,trees,positioning,arrows,chains}

\date{}

\begin{document}
\title{Parameter Estimation for the Single-Look $\mathcal{G}^0$ Distribution}
\author{D\'ebora Chan, Andrea Rey, Juliana Gambini and~Alejandro~C.~Frery,~\IEEEmembership{Senior Member,~IEEE}%
\thanks{D.\ Chan is with Universidad Tecnol\'ogica Nacional, Facultad Regional Buenos Aires, Ciudad Aut\'onoma de Buenos Aires, Argentina email: \texttt{mchan@frba.utn.edu.ar}}%
\thanks{A.\ Rey is with Universidad Tecnol\'ogica Nacional, Facultad Regional Buenos Aires, Ciudad Aut\'onoma de Buenos Aires, Argentina}%
\thanks{J.\ Gambini is with Departamento de Ingenier\'{\i}a Inform\'atica, Instituto Tecnol\'ogico de Buenos Aires, Buenos Aires, Argentina and Depto.\ de Ingenier\'{\i}a en Computaci\'on, Universidad Nacional de Tres de Febrero, Pcia.\ de Buenos Aires, Argentina}%
\thanks{A.\ C.\ Frery is with Universidade Federal de Alagoas, Brazil email: \texttt{acfrery@laccan.ufal.br}}%
}

\markboth{IEEE Geoscience and Remote Sensing Letters,~Vol.~X, No.~X, XXX}%
{Chan \MakeLowercase{\textit{et al.}}: Parameter Estimation $\mathcal G^0$}
\maketitle


\begin{abstract}
The statistical properties of Synthetic Aperture Radar (SAR) image texture reveals useful target characteristics.
It is well-known that these images are affected by speckle, and prone to contamination as double bounce and corner reflectors.
The $\mathcal{G}^0$ distribution is flexible enough to model different degrees of texture in speckled data. 	
It is indexed by three parameters:  $\alpha$, related to the texture, 
$\gamma$, a scale parameter, and $L$, the number of looks which is related to the signal-to-noise ratio. 
Quality estimation of $\alpha$ is essential due to its immediate interpretability. 
%
%
In this article, we compare the behavior of a number of parameter estimation techniques in the noisiest case, namely single look data.
We evaluate them using Monte Carlo methods for non-contaminated and contaminated data, considering convergence rate, bias, mean squared error (MSE) and computational cost.
The results are verified with simulated and actual SAR images.
\end{abstract}

\begin{keywords}
Speckle, Parameter Estimation, Contaminated Data
\end{keywords}

\section{Introduction}
\label{sec:1}

\IEEEPARstart{S}{AR} is an active sensing instrument able to measure the roughness, electrical properties and shape of the surface. 
It is widely used in environmental monitoring and evaluation of damages in natural catastrophes, among other applications.
However, automatic SAR image interpretation is difficult due to the presence of speckle noise, making statistical modeling necessary.

Many statistical models have been proposed for SAR image understanding.
Among them, the $\mathcal{G}^0$ distribution is able to characterize a large number of targets.
This model enhances the classical $\mathcal K$ distribution that fails to model extremely textured areas~\cite{mejail2003classification}. 
Based on the ability to model regions with different degrees of texture, it has been called the  ``Universal Model''~\cite{gao2010statistical}.
It is indexed by three parameters: $\alpha$, related to the texture, $\gamma$, a scale parameter, and $L$, the number of looks which is related to the signal-to-noise ratio. 
The last parameter may be known or estimated for the whole image, while the others describe local characteristics.

Due to the direct interpretation of  $\alpha$, that indicates the degree of texture of a region in a scene, precision and accuracy in its estimation are basilar for the development of procedures and algorithms that employ such estimates.
The aim of this paper is to assess the performance of several estimation methods for speckled data in the single look case ($L=1$), especially for data containing a corner reflector.

There are many estimation techniques, among them analogy methods, e.g. Moments (MOM)~\cite{gambini2015} and Log-Moments (LMOM)~\cite{1344160}, and Maximum Likelihood (MLE)~\cite{gambini2015}.  
Gambini et al.~\cite{gambini2015} proposed a non parametric method which consists in minimizing the Triangular Distance between the $\mathcal{G}^0$ density probability function and an empirical distribution of the data computed with asymmetric kernels.
This proposal is robust, but has high computational cost and that it fails in the single look case.
For this reason, we study estimation methods for $L=1$, the noisiest situation.

The most desirable estimator is MLE because of its asymptotic properties, even though it has problems in small samples regarding bias, convergence and robustness. 
Several attempts have been made to reduce MLE bias using analytic~\cite{giles2016bias,da2008improved} and bootstrap methods~\cite{vasconcellos2005improving}. 
Other efforts have been oriented towards correcting its tendency to diverge with small samples~\cite{FreryCribariSouza:JASP:04}.

Robustness is a desired property.
Among the possible deviations from the ideal situation of iid deviates, extreme values are frequent in SAR imagery due to, for instance, corner reflectors and other sources of double bounce.
Among the robust proposals, M-estimators proved to be reliable in the presence of corner reflectors~\cite{BustosFreryLucini:Mestimators:2001,bustos2002m}.
Robust AM-estimators~\cite{allende2006m} perform similarly as MLE. 
For certain contamination schemes the AM-estimator, built with an asymmetric influence function, outperforms MLE.

For the single look case, the $\mathcal{G}^0$ distribution is a Generalized Pareto type II distribution~\cite{chan2016texture}. 
This law has been verified and studied in many fields because of its flexibility to model different phenomena. 
We take advantage of this fact, and check estimators for this distribution.

We assess parameter estimation techniques  under the single look $\mathcal G^0$ model according to their computational cost, convergence rate, bias and mean squared error for data with and without contamination using Monte Carlo.
We compare the performance of threshold selection techniques for parameter estimation, and we then apply these methods to actual data with corner reflectors.

This article  unfolds as follows.  
Section~\ref{sec:2} recalls $\mathcal{G}^0$ distribution properties. 
Section~\ref{sec:3} introduces the selected estimators.  
Section~\ref{sec:4} discusses the results obtained with simulations for contaminated and non contaminated data. 
Section~\ref{sec:5} shows applications to actual data. 
Finally, Section~\ref{sec:6} discusses remarks and presents conclusions.

\section{Single look $\mathcal{G}^0$ model}
\label{sec:2}

The density probability function of the $\mathcal{G}^0$ distribution for intensity data is given by
\begin{equation}
f_{\mathcal{G}^{0}}( z) =\frac{L^{L}\Gamma ( L-\alpha
	) }{\gamma ^{\alpha }\Gamma ( -\alpha ) \Gamma (L) }\cdot  
\frac{z^{L-1}}{( \gamma +zL) ^{L-\alpha }},
\label{ec_dens_gI0}
\end{equation}
where $-\alpha,\gamma ,z>0$ and $L\geq 1$. 
We are interested in the noisiest case, which corresponds to $L=1$, called the single look case. 
The probability density function becomes:
\begin{equation}
	f_{\mathcal{G}^{0}}( z) =\frac{-\alpha}\gamma \left(\frac{z}\gamma +1 \right)^{\alpha-1} .
\label{gi0 density}
\end{equation}
The $r$-order moments for the single look case are 
\begin{equation}
\text{E}(Z^r) =\gamma^r\frac{\Gamma ( -\alpha-r )}{ \Gamma (-\alpha) }
{\Gamma (1+r )}
\label{moments_gI0}
\end{equation}
if $\alpha<-r$ and infinite otherwise.

The Generalized Pareto Type II Distribution (GPD), $GP_{\rm{II}}(\mu, \sigma, \beta)$ has
probability density function given by:
\begin{equation}
f_{GP_{\rm{II}}}(z) =\frac{\beta}{\sigma}\Big(1+\frac{z-\mu}{\sigma}\Big)^{-\beta-1}  ,
\label{pareto density} 
\end{equation}
so the $\mathcal{G}^0(\alpha, \gamma, 1)$ distribution is a particular case of this distribution for 
$\mu =0$, $\sigma=\gamma$ and $\beta=-\alpha$.
We take advantage of this observation.

Every distribution can be characterized by its moments and by its Probability-Weighted Moments (PWM)~\cite{greenwood1979probability}, defined as

\begin{equation}
M_{p,r,s}= \text{E} \left[Z^p(F(Z))^r(1-F(Z))^s \right] ,
\end{equation}
where  $F(z)=\Pr(Z\leq z)$ is the cumulative distribution function (CDF) and $p, r, s \in \mathbb R$.
In the special case $p = 1$ and $r= 0$ under the $\mathcal{G}^0$ distribution, one has
\begin{equation}
M_{1,0,s}=\frac{\gamma}{(s+1)\left[1+\alpha(s+1) \right] }.
\end{equation}

Gambini et al.~\cite{gambini2015} proved several properties of the $\mathcal{G}^0$ distribution, among them that it varies slowly at infinite, its heavytailedness with tail index $1-\alpha$, and that it is prone to produce outliers~\cite{rojo2013heavy}. 

\section{Parameter Estimation Methods: State of Art}
\label{sec:3}

In this section we review the following estimation methods: 
Maximum Likelihood,
Penalized Maximum Likelihood,
Moments (based on the first and second),
Penalized Weighted Moments,
Likelihood Moments,
Median,
Minimum Power Diversity Divergence,
Maximum Goodness of Fit, and their variants.

Given the sample  $\bm{z}=(z_1,\dots, z_n)$ of independent and identically distributed random variables with common distribution $\mathcal{G}^0(\alpha,\gamma,1)$ with $(\alpha,\gamma) \in \Theta = \mathbbm{R}_- \times \mathbbm{R}_+ $, a maximum likelihood estimator $(\widehat\alpha,\widehat\gamma)$ satisfies 
\begin{eqnarray}
n[\Psi^0(-\widehat{\alpha})-\Psi^0(1-\widehat{\alpha})]+\sum_{i=1}^n \ln\frac{\widehat{\gamma}+z_i^2}{\widehat{\gamma}}=0\\
\frac{n\widehat{\alpha}}{\widehat{\gamma}}+(1-\widehat{\alpha})\sum_{i=1}^n(\widehat{\gamma}+z_i^2)^{-1}=0,
\end{eqnarray}
where $\Psi^0(t) = {d\ln\Gamma(t)}/{dt}$ is the digamma function. 

Among others, 
Grimshaw~\cite{grimshaw1993computing} studied the properties of the MLE under the GPD model.
It is asymptotically consistent, efficient and normal, but in many cases it has not an explicit solution and it diverges for small samples.
Then, numerical techniques are required.

With the purpose of reaching convergence for small samples, Coles and Dixon~\cite{coles1999likelihood} proposed maximizing the log-likelihood minus a penalization function, for instance 
$\exp\{ {\eta}/({\eta-1}) \}$,
$\eta\in(0,1)$. 
For large sample sizes, MPLE inherits optimal properties from MLE, avoiding MLE limitations in small ones.

From~\eqref{moments_gI0}, the MOM estimator is
\[\widehat{\alpha}_{\text{MOM}}=\frac{2s^2}{\overline{\bm{z}}^2-s^2}  \text{ and } 
\widehat{\gamma}_{\text{MOM}}=\frac{\overline{\bm{z}}(\overline{\bm{z}}^2+s^2)}{\overline{\bm{z}}^2-s^2}.
\]	
where $\overline{\bm{z}}$ and $s^2$ denote sample mean and variance, respectively.
Hosking and Wallis~\cite{hosking1987parameter} discussed some properties of MOM and Probability Weighted Moment (PWM) estimators for the GPD distribution. 
They compared the performace of MOM, MLE and PWM and observed that MOM is asymptotically normal but also that it frequently does not  converge and sensitive to outliers.
They  showed that PWM is an alternative to ordinary moments with advantages for small sample sizes but with low asymptotic efficiency.
The PWM expression is 
\[\widehat{\alpha}_{\text{PWM}}=\frac{\overline{\bm{z}}^2-2\theta}{4\theta-\overline{\bm{z}}}\qquad \text{and} \qquad
\widehat{\gamma}_{\text{PWM}}=\frac{2\theta \overline{\bm{z}}}{\overline{\bm{z}}-4\theta}.
\]	
 where $\theta={n}^{-1}\sum_{i=1}^n \frac{n-i}{n-1} \bm{z}_i$.


Zhang~\cite{zhang2007likelihood} proposed the Likelihood Moment Estimator (LME) by combining  MLE and MOM techniques.  
The solution always exists, is efficient and asymptotically normal.

Peng and Welsh~\cite{peng2001robust} proposed the Median Estimator (MED), as a robust alternative, by solving an equation that equals the sample and population likelihood score medians.
It has robustness because of its bounded influence function, and it is asymptotically normal, but in many cases it does not have good performance~\cite{juarez2004robust} being, thus, only advisable under the presence of outliers.
 
The Minimum Power Density Divergence (MDPD) estimator is another robust alternative.
It has bounded influence function and is indexed by a positive constant $\omega$ which controls the trade-off between efficiency and robustness: 
MDPD becomes MLE when $\omega=0$. 
Juarez and Schucany~\cite{juarez2004robust} proved that MLE is the most efficient for non contaminated data but MDPD outperforms it under contamination.

Luce\~no~\cite{luceno2006fitting} proposed maximizing the Maximum Goodness of Fit (MGF) estimator.  
They showed its consistency and asymptotic efficiency. 
This method allows the use of several goodness of fit statistics; in this work we evaluate the Kolmogorov-Smirnov test (KS), Cramer Von Mises (CM), Anderson Darling (AD), Right tail AD (ADR), Left tail AD (ADL), Second order Right tail AD (AD2R), Second order Left tail AD (AD2L) and Second order AD (AD2). 

\begin{figure}[hbt]
	\centering
	\scalebox{.5}{
	\begin{tikzpicture}[node distance=1.5cm]  
	\tikzset{
		mynode/.style={rectangle,rounded corners,draw=black, top color=white, bottom color=blue!50,very thick, inner sep=1em, minimum size=1em, text centered},
		myarrow/.style={->, >=latex', shorten >=1pt, thick},
		mylabel/.style={text width=4em, text centered} 
	}  
	\node[mynode] (rob) {Robust};  
	\node[below=2cm of rob] (dummy) {}; 
	\node[mynode, left=of dummy] (AN) {Asymp. Norm.};  
	\node[mynode, right=of dummy] (AEN) {Asymp. Eff. Norm.};
	\node[mylabel, below left=of rob] (label1) {Yes};  
	\node[mylabel, below right=of rob] (label2) {No};
	\node[mylabel, below left=of AEN] (label1) {High};  
	\node[mylabel, below right=of AEN] (label2) {Low};
	
	\draw[myarrow] (rob.south) -- ++(-.5,0) -- ++(0,-1) -|  (AN.north);	
	\draw[myarrow] (rob.south) -- ++(.5,0) -- ++(0,-1) -|  (AEN.north);
	
	\node[below=0.1cm of AN] (primer) {};
	\node[mynode, below=of primer] (CC) {Comp. Cost};
	\draw[myarrow] (AN.south) to (CC.north);  
	
	\node[below=2.2cm of AEN] (segundo) {}; 
	\node[mynode, left=of segundo] (NES) {Non exp. sol.};  
	\node[mynode, right=of segundo] (ES) {Exp. sol.};

	\draw[myarrow] (AEN.south) -- ++(-.5,0) -- ++(0,-1) -|  (NES.north);	
	\draw[myarrow] (AEN.south) -- ++(.5,0) -- ++(0,-1) -|  (ES.north);
	
	\node[below=2cm of CC] (izq) {}; 
	\node[mynode, left =of izq] (med) {MED};  
	\node[mynode, right=of izq] (mdpd) {MDPD-MGF};
	\node[mylabel, below left=of CC] (label1) {High};  
	\node[mylabel, below right=of CC] (label2) {\;\; \; \;Low};
	\draw[myarrow] (CC.south) -- ++(-.5,0) -- ++(0,-1) -|  (med.north);	
	\draw[myarrow] (CC.south) -- ++(.5,0) -- ++(0,-1) -|  (mdpd.north);
	
	\node[below=1cm of NES] (medio) {}; 
	\node[mynode, right=of medio] (lme) {LME-MLE-MPLE};  
	\draw[myarrow] (NES.south) to (lme.west);
	
	\node[below=0.2cm of ES] (der) {}; 
	\node[mynode, below=of der] (mom) {MOM};  
	\node[mynode, below = of mom] (pwm) {PWM};
	\draw[myarrow] (ES.south) to (mom.north);
	\draw[->, >=latex', shorten >=2pt, shorten <=2pt, bend left=45, thick] 
	(ES.south) to node[auto, swap] {}(pwm.east); 
	
\end{tikzpicture}}
\caption{Parameter estimation methods characteristics.}
\label{Estimatorcharacteristcsscheme} 
\end{figure}
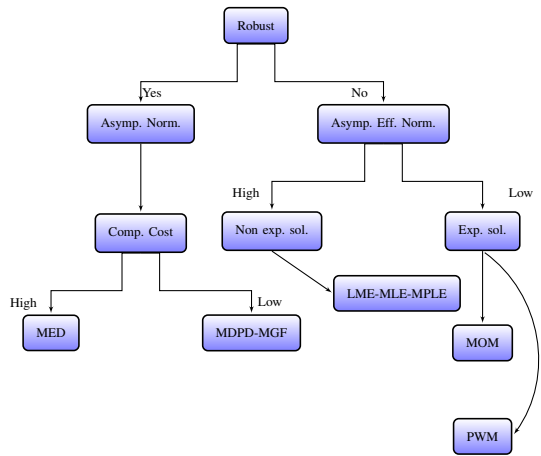

Figure~\ref{Estimatorcharacteristcsscheme} summarizes the main characteristics of the considered parameter estimation methods, where ``Asymp.\ Norm", ``Asymp.\ Eff.\ Norm." and ``Exp.\ Sol." mean Asymptotic Normality, Asymptotic Efficient Normality and Explicit Solution, respectively. 

The Generalized Pareto Distribution (GPD) is the limiting law of normalized excesses over a thereshold~\cite{pickands1975statistical}.
Thus, the choice of threshold is crucial for accurate estimation. 
We assessed the performance of the aforementioned estimators with the following threshold selection methods: 
(a)~$u_0$ which considers the whole sample, 
(b)~$u_{q_{10}}$ which uses the \SI{90}{\percent} largest values, 
(c)~$u_{q_{20}}$ which considers the \SI{80}{\percent} largest values, 
(d)~$u_{\text{Hill}}$ which is based on the Hill plot, and 
(e)~$u_{\text{AD}}$ which is an automated threshold selection based on the $p$-values of the AD goodness of fit test.

In order to decide the most suitable threshold for each estimator, we generated $1000$ samples of sizes $25$, $49$ and $81$, for all combinations of $\alpha \in \{ -8, -5, -2\}$ and $\gamma \in \{ 0.1, 1 ,10 \}$.
We conclude that $u_{q_{10}}$ is the best threshold for $n=49$, for MDPD and MLE.
In the other cases, $u_0$ is better.

\section{Simulation Experiments and Results}
\label{sec:4}
In this section we compare the following estimation methods: ADR, MDPD, MPLE, LME, MLE and PWMB by their mean squared error (MSE), convergence rate, bias and computational time, for non contaminated and contaminated data.

\subsection{Non contaminated data}
\label{subsec:3-1}

We consider $1000$ samples from the $\mathcal{G}^0(\alpha, \gamma,1)$ distribution, of sizes  $25, 49, 81, 121, 500$ without contamination combining the parameter values $\alpha\in \left\lbrace -8,-5,-2\right\rbrace $ and $\gamma\in \left\lbrace 0.1, 1, 10, 100 \right\rbrace$.
Samples were obtained following the guidelines presented in Ref.~\cite{SamplingfromtheGI0Distribution2018}.

We compute the estimates using the six methods mentioned above, and then we assess them by analyzing  computational cost, convergence rate, bias and mean squared error.

\begin{figure}[hbt]
	\centering
	\includegraphics[width=\linewidth]{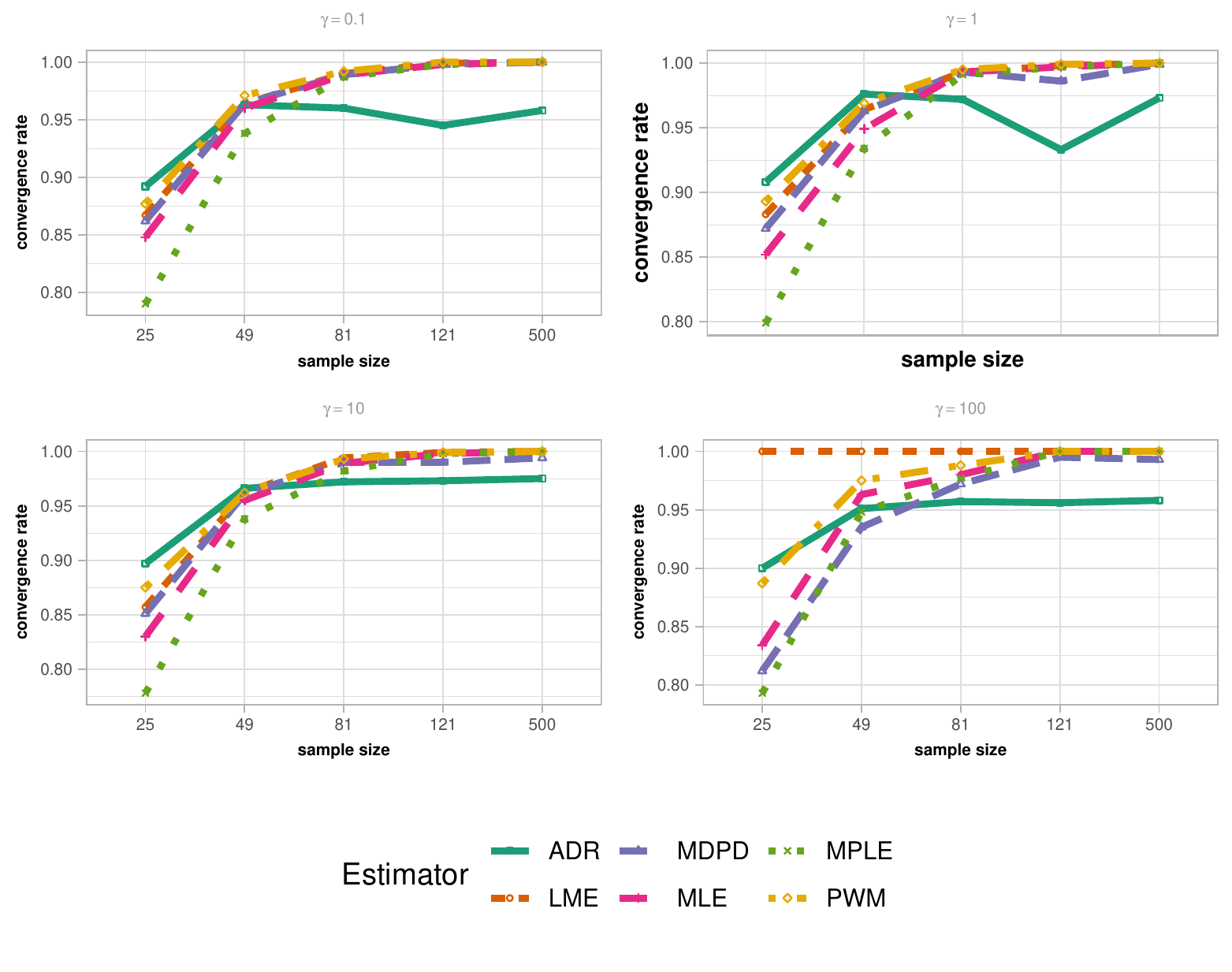}
	\caption{Convergence rate for $\alpha=-2$.}
	\label{fig:convnocont2}
\end{figure}

The plots in Figure~\ref{fig:convnocont2} show the convergence rate for $\alpha =-2$. 
For small samples, the best convergence rate was observed for ADR, but its performance is not acceptable for large samples. 
Except for MPLE, the rest of the considered estimators reach good convergence rate in large samples and an adequate one in small samples.

\begin{figure*}[hbt]
	\centering
\subfigure[Biases for $\alpha=-8$.\label{fig:biasnocont8}]{\includegraphics[width=.48\linewidth]{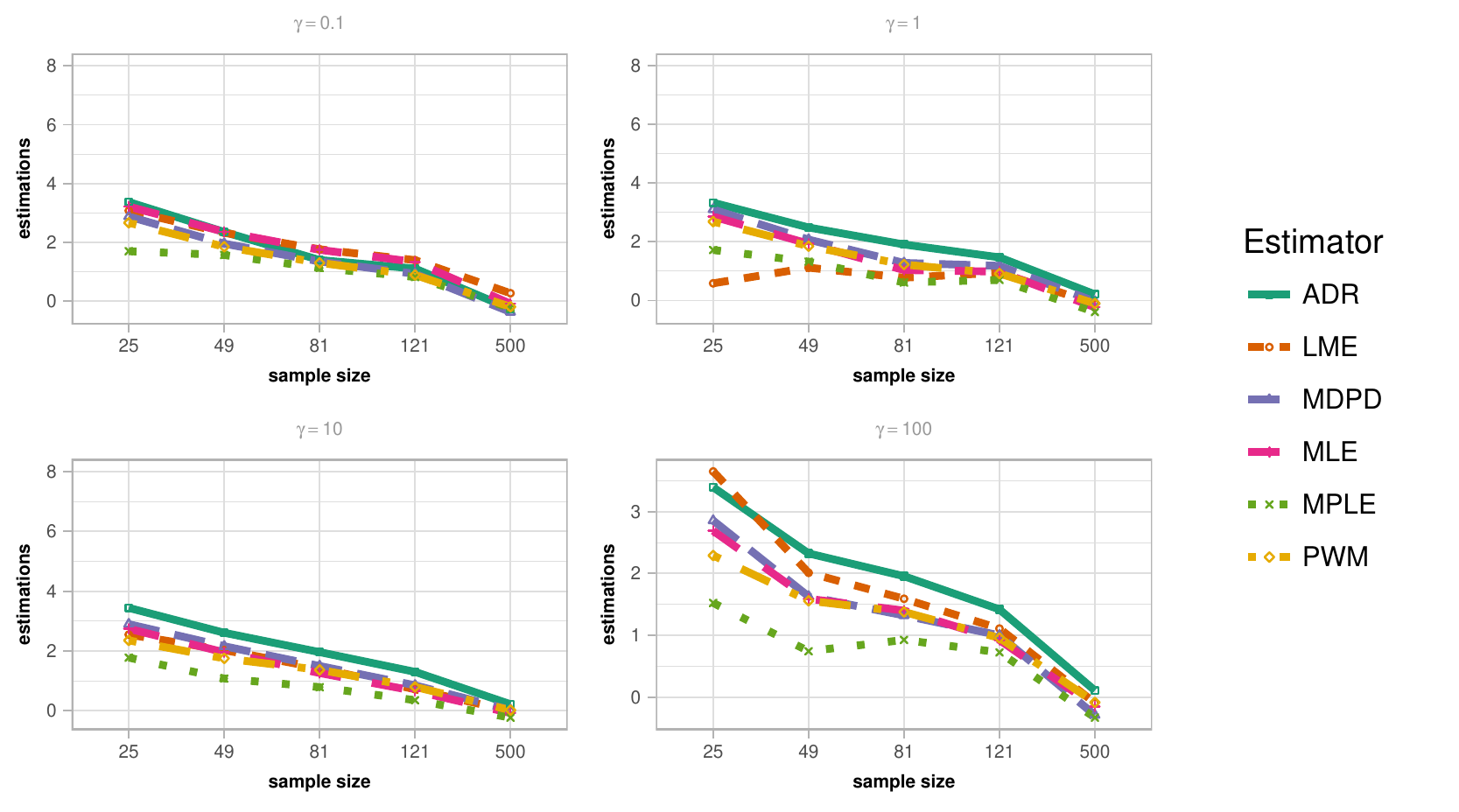}}
\subfigure[Mean Square error for $\alpha=-5$\label{fig:msenocont5}]{\includegraphics[width=.48\linewidth]{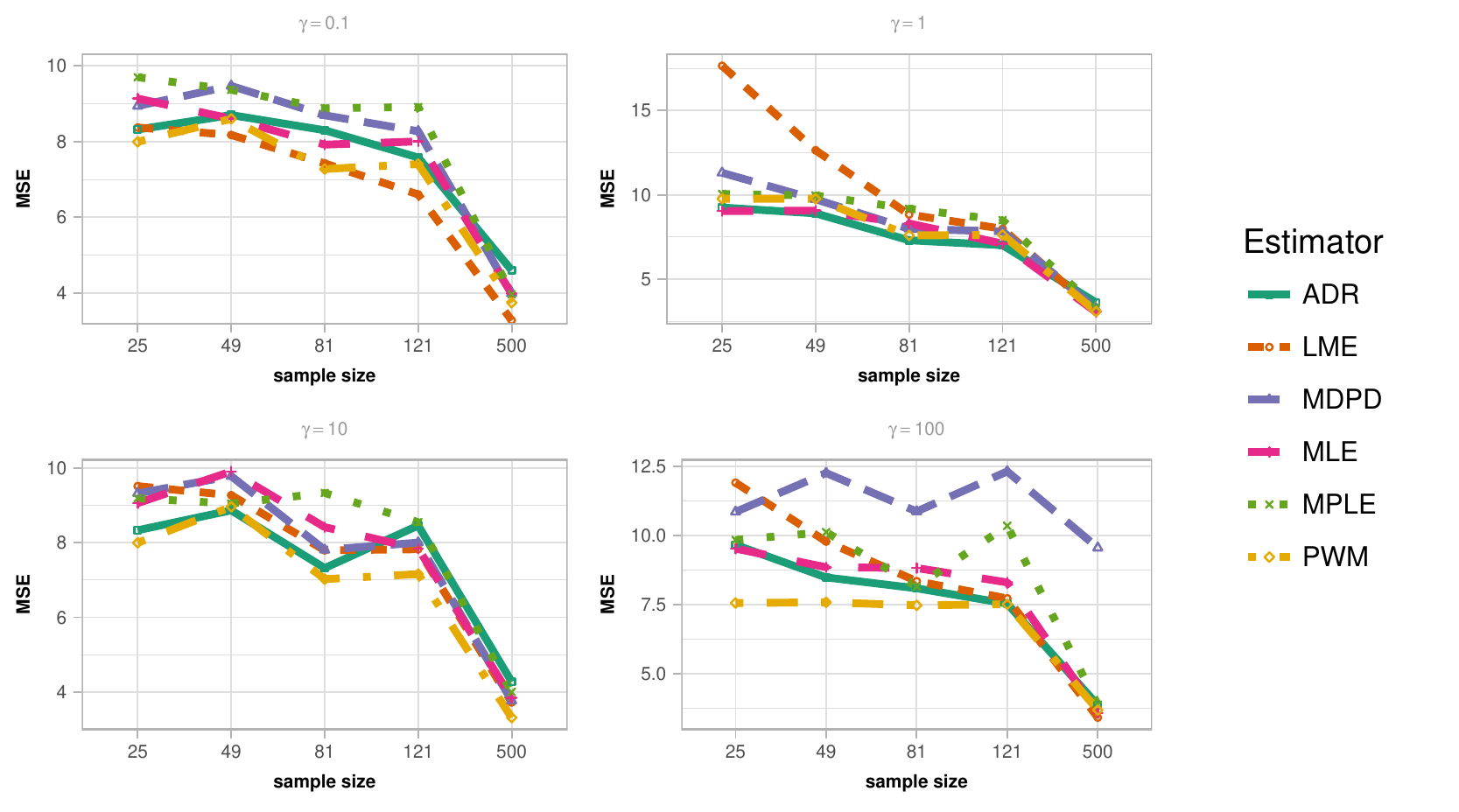}}
	\caption{Bias and Mean Square Error.}
	\label{fig:MeanBias}
\end{figure*}

Figure~\ref{fig:biasnocont8} shows the bias for $\alpha = -8$ (blue line). 
We note that the bias is greater than $2$ in small sample sizes for all estimators. 
It can be seen that the best accuracy was achieved by MPLE and in some cases by LME, meanwhile  all estimators achieve a good accuracy in sample sizes larger than $121$.
Figure~\ref{fig:msenocont5} illustrates the MSE for $\alpha =-5$. 
The performance is similar for all the candidates, but we point that MDPD and ADR have erratic performance in some cases and that LME and PWM mark superiority for almost all cases.

\begin{figure}[hbt]
	\centering
	\includegraphics[width=.8\linewidth]{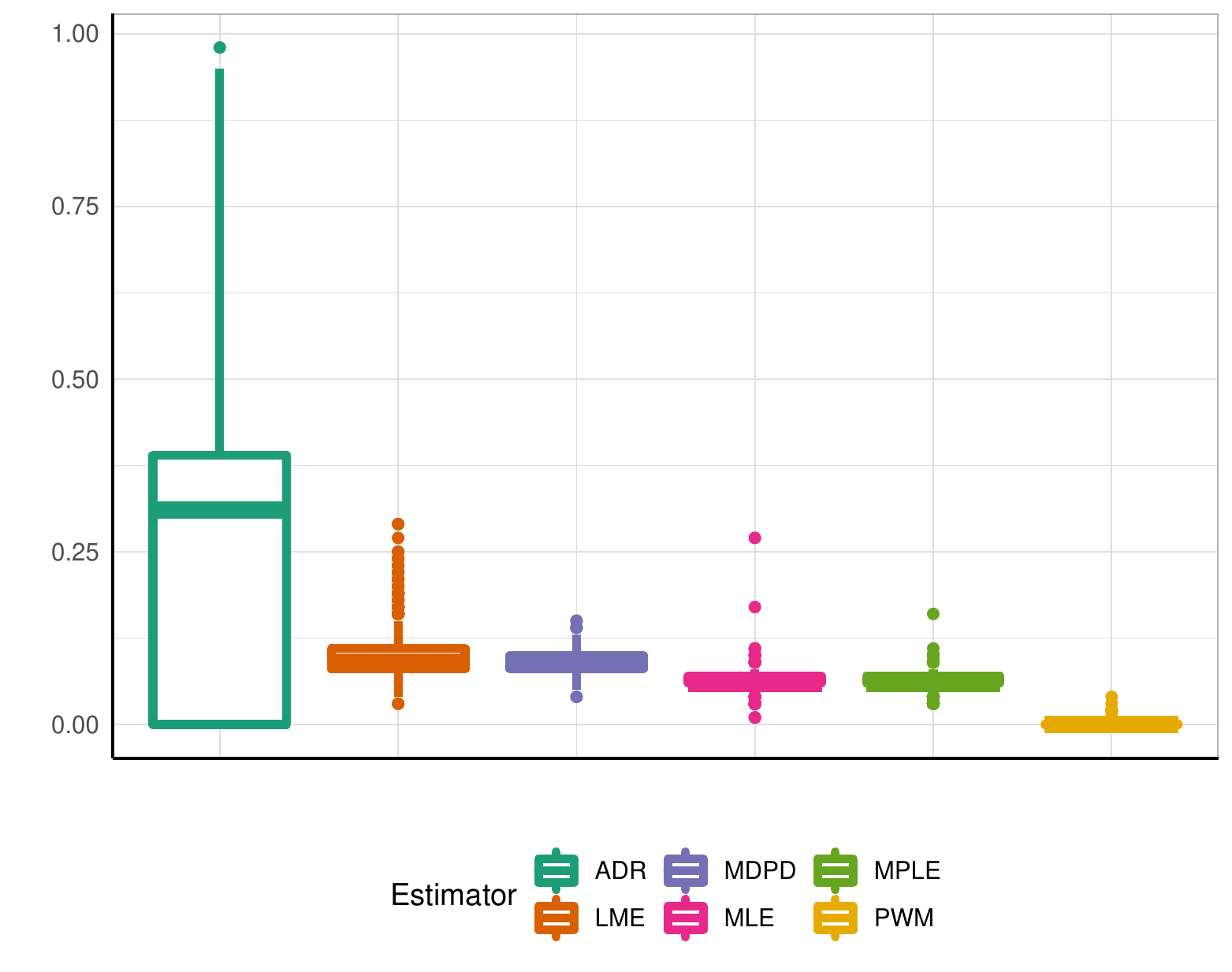}
	\caption{Processing time in milliseconds by each method.}
	\label{fig:times}
\end{figure}

Figure~\ref{fig:times} shows the time consumed in milliseconds by each method for $1000$ samples of size $500$ for all parameter combinations. 
The multiple comparisons of Tukey HSD test point out that PWM method is better than the others with respect to the computational time.

\subsection{Contaminated Data}
Extremely high values may appear in SAR imagery due to the double bounce effect, e.g. in the presence of corner reflectors.
Such outliers may cause gross errors in the parameter estimation of the background.
We generated contaminated random samples in order to assess the estimators under this kind of contamination. 

We describe the occurrence of contamination with a Bernoulli random variable $B$ with probability of success $0<\epsilon\ll 1$.
Let $C \in \mathbb R_+$ be a large value.
The random variable $Z=BC+(1-B)W$ is our model for the return of the background  $W \sim \mathcal{G}^0(\alpha,\gamma,1)$ under contamination.

As a way of measuring the impact of this contamination, we constructed stylized empirical influence functions (SEIFs)~\cite{allende2006m} for samples of sizes $n \in \{25, 49, 81,121\}$, for each estimator considering all parameter combinations.

\begin{figure}[hbt]
	\centering
	\includegraphics[width=\linewidth]{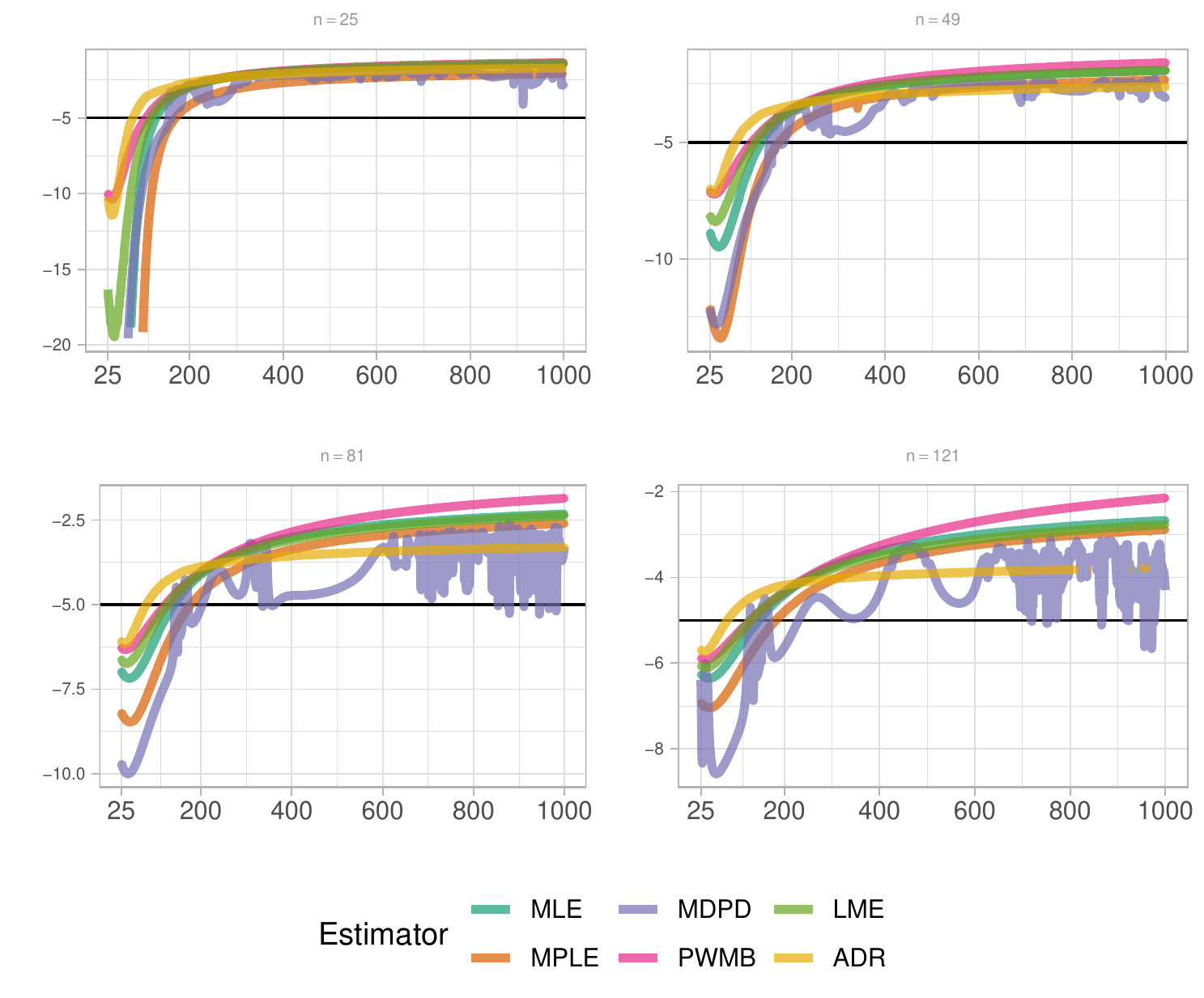}
	\caption{Stylized empiric influence functions, $\alpha=-5,\gamma=100$.}
	\label{fig:influence}
\end{figure}

Figure~\ref{fig:influence} shows such functions for $\alpha =-5$ and $\gamma = 100$.
With this, the expected value of the uncontaminated background is $25$;
the abscissas span from $25$ to $1000$.

The smaller the SEIF is, the better the performance of the estimator is before contamination.
MPLE and LME outperform the other estimators, except for the robust MDPD, specially for large contamination values.

\section{Actual SAR Data}
\label{sec:5}

\begin{figure}[hbt]
	\centering
		\subfigure[Single-look E-SAR image with a corner reflector.]{\includegraphics[width=.7\linewidth]{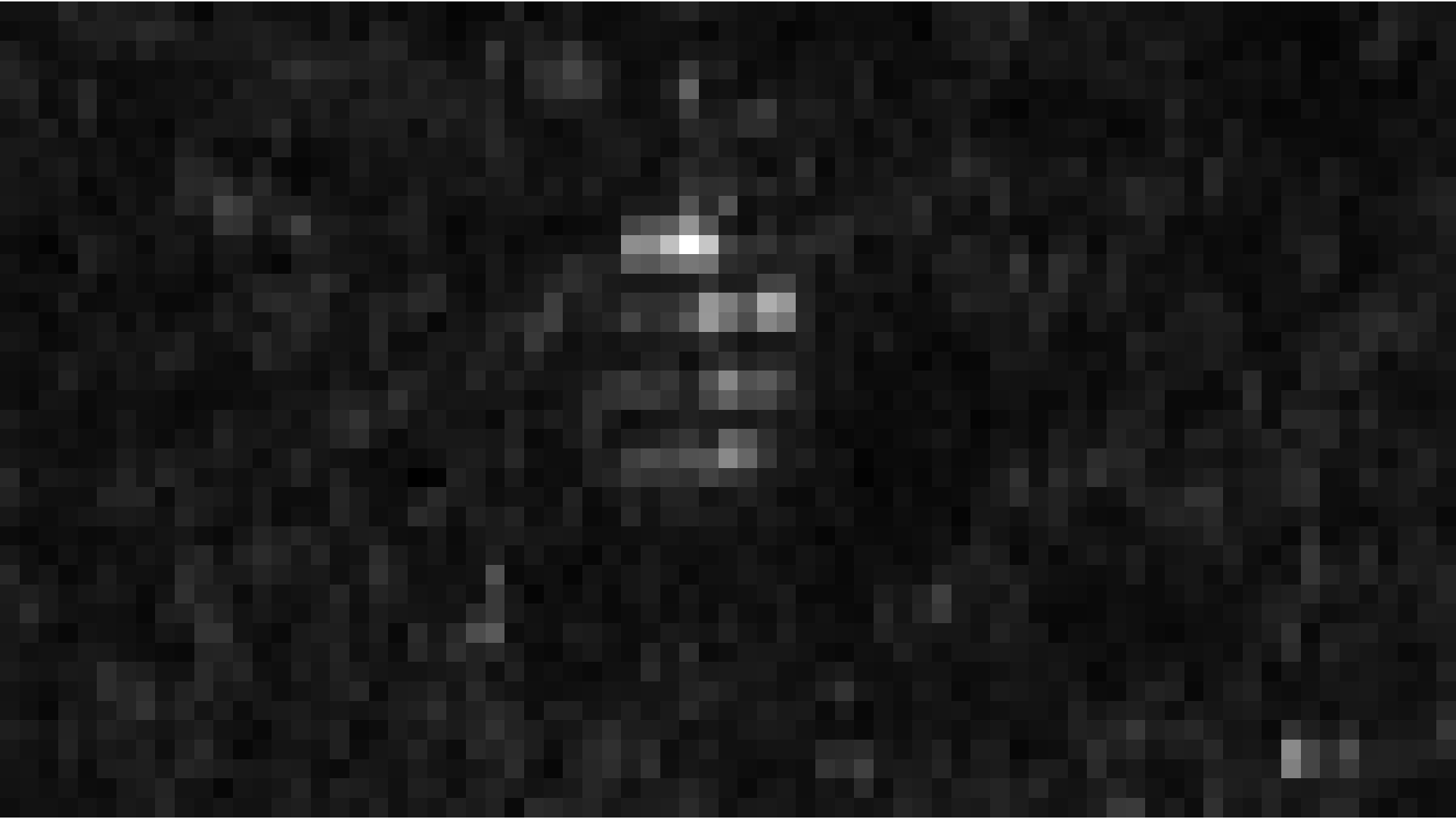}\label{fig:samplesa}}
		\subfigure[Regions of interest used to compute the estimates of the texture parameter. ]{\includegraphics[width=.7\linewidth]{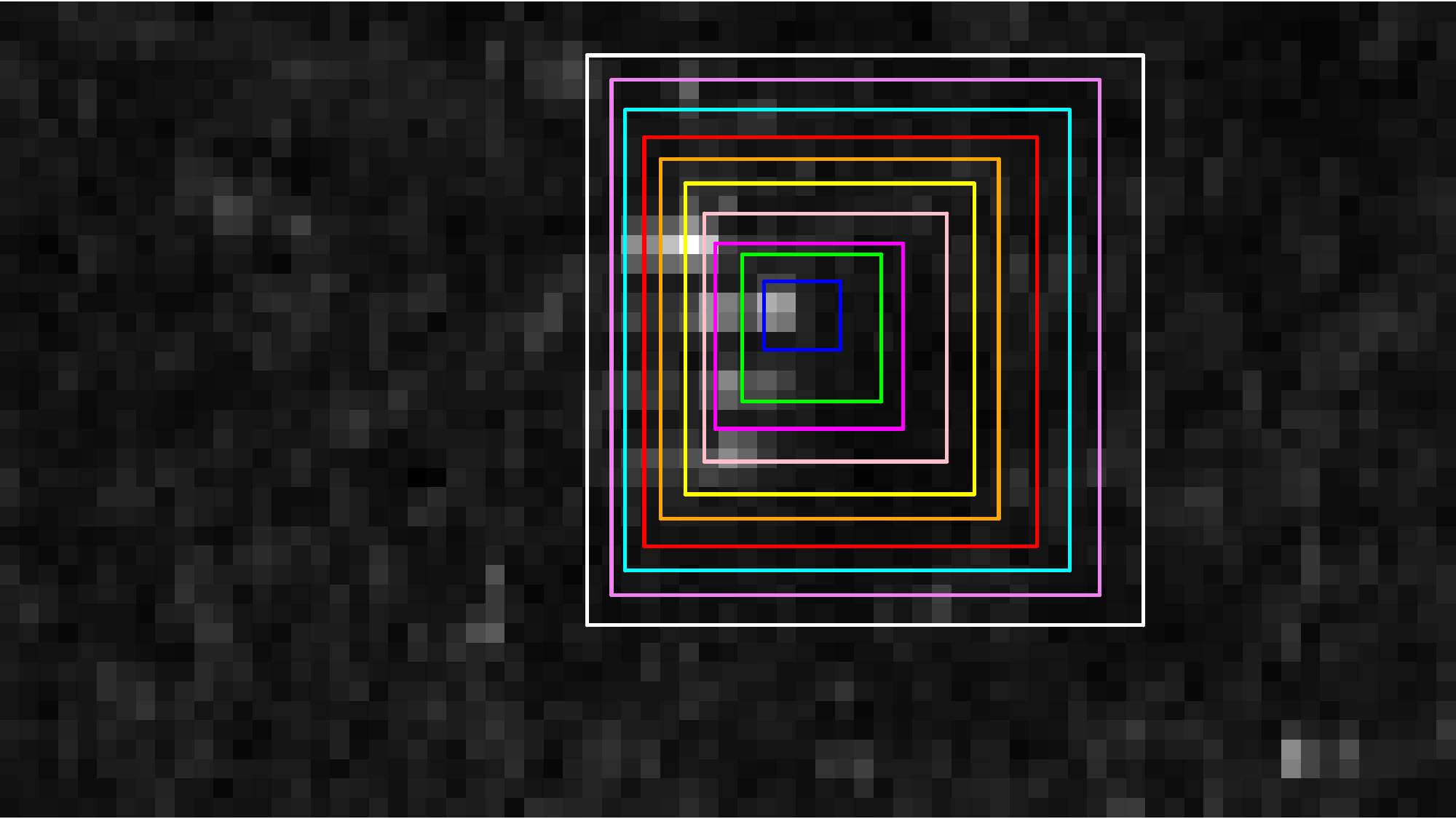}\label{fig:samplesb}}
	\caption{Single-look E-SAR image with a corner reflector, used to estimate the $\alpha$-parameter (\ref{fig:samplesa}). Ten Regions of interest of different sizes used to estimate the texture parameter (\ref{fig:samplesb}). }
	\label{fig:samples}
\end{figure}

\begin{figure}[hbt]
	\centering
	\includegraphics[width=.95\linewidth]{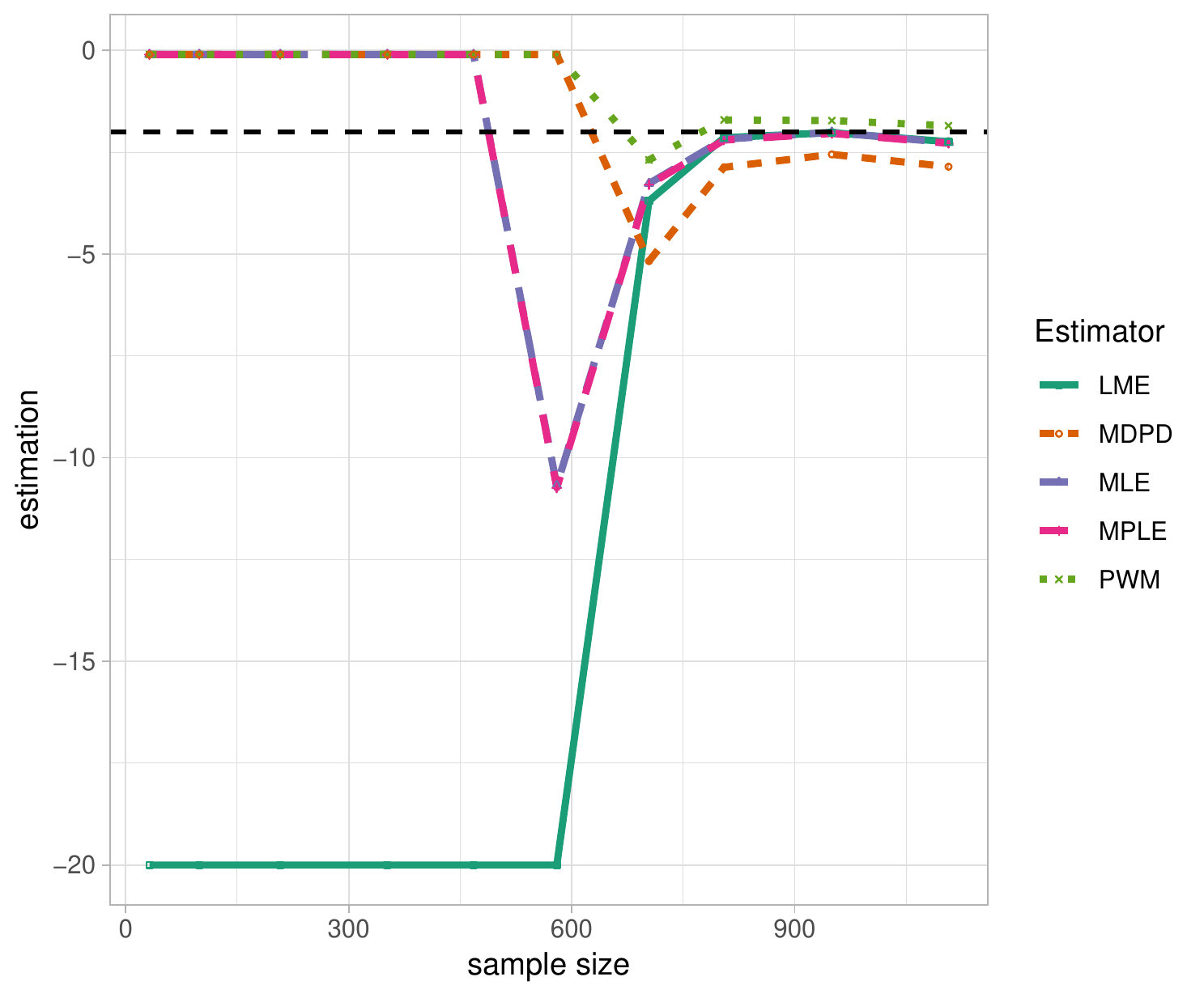}
	\caption{Texture estimates using the samples from Figure~\ref{fig:samplesb}.}
	\label{fig:imageest}
\end{figure}

Fig.~\ref{fig:samplesa} shows the intensity single-look L-band HH polarization E-SAR image with a corner reflector used to compute the estimates.
Fig.~\ref{fig:samplesa} shows the regions used for estimating the texture.
The estimates are presented in Fig.~\ref{fig:imageest}, and their values along with the sample sizes are showed in Table~\ref{TablaDeEstimacionesDeCorner}.
The black dashed line is the true value of $\alpha$, as informed in~\cite{gambini2015}.
For samples of size greater than $600$, estimations are near the true value for all the techniques, although MDPD is the most biased. 

\begin{table}[hbt]
	\centering
	\caption{Texture parameter  estimations for the regions of  Figure~\ref{fig:samplesb}.}
	\label{TablaDeEstimacionesDeCorner}
	\begin{tabular}{@{}rrrrrr@{}}
		\toprule
			   Sample size   &    MLE   &   MPLE     &   LME  &   PWM  &    MDPD \\
			   \midrule
			  $36$ & $-0.10$  & $-0.10$ & $-20.00$ & $-0.10$ & $-0.10$ \\
		      $63$ & $-1.62$ & $-1.93$ & $-1.55$ & $-1.56$ & $-1.98$ \\
			  $121$ & $-2.38$ & $-2.52$ &  $-2.22$ & $-1.92$ & $-3.51$ \\
			  $168$ & $-2.85$ & $-2.96$ &  $-3.18$ & $-2.28$ & $-4.69$ \\
			  $270$ & $-3.44$ & $-3.51$ & $-4.17$ & $-2.73$ & $-6.57$  \\
			  $396$ & $-3.87$ & $-3.92$ & $-4.35$ & $-3.37$ & $-7.75$ \\
			  $468$ & $-3.58$ & $-3.62$  &  $-3.93$ & $-3.05$ & $-6.11$ \\
			  $540$ & $-3.59$ & $-3.63$ &  $-4.22$ & $-3.00$ &  $-6.43$ \\
			  $665$ & $-2.33$ & $-2.36$ &  $-2.39$ & $-1.74$ & $-3.38$ \\
			 $740$ & $-2.29$ &  $-2.32$ &  $-2.32$ & $-1.74$ & $-3.16$ \\
		\bottomrule
\end{tabular}
\end{table}

\section{Final Remarks}
\label{sec:6}

In this article, we compared six parameter estimation methods with and without contamination.  We evaluate MSE, convergence rate, bias and computational time.

We concluded that
\begin{itemize}
 \item ADR method has the highest computational cost and the lowest convergence rate, for large samples.
 \item ADR has good convergence rate for small samples. 
 \item For ADR method, MSE and bias are acceptable for samples larger than $121$ in non contaminated data. 
 \item ADR estimation is precise for samples larger than $100$ in the presence of outliers. 
 \item The most accurate estimation under contamination is MDPD.
 \item MPLE and LME methods outperform the others techniques under contamination data,  except for the robust MDPD, specially for large outliers.
 \item MDPD is the most accurate estimator for actual small samples.
\end{itemize}

\appendix
Simulations were performed using the \texttt{R} language and environment for statistical computing version~3.3~\cite{Rmanual}, in a computer with processor Intel\textcopyright \ Core\texttrademark, i7-4790K CPU \SI{4}{\giga\hertz}, \SI{16}{\giga\byte} RAM, System Type \SI{64}{\bit} operating system. 

\bibliographystyle{IEEEtran}
\bibliography{References}
\end{document}